\documentclass[aps,prl,twocolumn,showpacs,preprintnumbers,amsmath,amssymb]{revtex4-1}

 \def\ep{{\epsilon}}

 \def\frac#1#2{{#1\over #2}}

 \def\s{\sqrt}

\def\be{\begin{equation}}
\def\ee{\end{equation}}
\def\ba{\begin{eqnarray}}
\def\ea{\end{eqnarray}}

 \def\de{\partial}

 \def\ti{\tilde}

 \def\ddd{\cdot\cdot\cdot}
 \def\no{\nonumber \\}

 \def\la{\langle}
 \def\lb{\rangle}
 \def\ep{\epsilon}

\usepackage{color}
\usepackage{graphicx}
\usepackage{dcolumn}
\usepackage{bm}
\usepackage{hyperref}

\begin{document}

\title{Holographic Entanglement of Purification}
YITP-17-89 ; IPMU17-0115
\author{Tadashi Takayanagi$^{a,b}$ and Koji Umemoto$^{a}$}

\affiliation{$^a$Center for Gravitational Physics, \\
Yukawa Institute for Theoretical Physics,
Kyoto University, \\
Kitashirakawa Oiwakecho, Sakyo-ku, Kyoto 606-8502, Japan}

\affiliation{$^{b}$Kavli Institute for the Physics and Mathematics
 of the Universe,\\
University of Tokyo, Kashiwa, Chiba 277-8582, Japan}

\date{\today}

\begin{abstract}
We study properties of the minimal cross section of entanglement wedge which connects two disjoint subsystems in holography. In particular we focus on various inequalities
which are satisfied by this quantity. They suggest that it is a holographic counterpart of the quantity called entanglement of purification, which measures a bipartite correlation in a given mixed state. We give a heuristic argument which supports this identification based on a tensor network interpretation of holography. This implies that the entanglement of purification satisfies the strong superadditivity for holographic conformal field theories.
\end{abstract}

\maketitle


\section{1. Introduction}

The entanglement entropy is a unique quantity which nicely characterizes quantum entanglement between two subsystems $A$ and $B$ for a given pure state. In the light of AdS/CFT \cite{Ma}, the entanglement entropy  has a simple holographic counterpart given by the area of minimal surface \cite{RT,HRT}. This gives a close relationship between spacetime geometry and quantum entanglement \cite{Ra,Swingle,MT,HAPPY,HQ,FH,MTW,CMNTW}.

One of the most important properties of entanglement entropy, called strong subadditivity, was derived geometrically using the holographic entanglement entropy in \cite{HeTa}. Moreover, a stronger inequality called monogamy of mutual information was derived in \cite{HHM} and this gives an interesting characterization of quantum states dual to a classical gravity background via the holography (see also \cite{H}). A large class of such entropic inequalities for holographic states has been found in \cite{Bao}.

On the other hand, for mixed states, many quantities which measure quantum or classical correlations
(including quantum entanglement) between two subsystems, called $A$ and $B$ below, have been known in quantum information theory \cite{HHHH,Book} (for a brief summary, refer also to appendix A of the present paper). We know essentially nothing about their holographic interpretations. Only one exception is the mutual information $I(A:B)=S(\rho_{A})+S(\rho_{B})-S(\rho_{AB})$ (here $AB\equiv A\cup B)$.  However,
since this quantity is just a linear combination of entanglement entropy, we cannot regard it as a genuinely new quantity from the view point of either holographic or quantum information theory. This motivates us to explore an independent quantity which measures a correlation between two subsystems for a mixed state and has a clear holographic interpretation.

If we have in mind holographic computations based on the AdS/CFT correspondence, there is another interesting candidate which measures correlation between two disjoint subsystems $A$ and $B$. Consider a static example of AdS/CFT whose boundary consists of the subsystem $A$, $B$ and the complement of $AB$ at a fixed time. The bulk region dual to a reduced density matrix $\rho_{AB}$ is called the entanglement wedge \cite{EW1,EW2,EW3}
(more precisely the restriction of entanglement wedge on the canonical time slice), which we will write $M_{AB}$.
The candidate which we would like to study in this paper is the minimal cross section of the entanglement wedge, which separates the wedge into two parts: the one includes $A$ and the other one $B$. We write this as $E_W(\rho_{AB})$ and call it entanglement wedge cross section. This quantity measures a certain correlation between two subsystems. The main purpose of this paper is to explore its properties and interpretation in conformal field theories (CFTs) by employing quantum information theoretic considerations.

\section{2. Holographic Entanglement Entropy}

Let us start with the holographic computation of entanglement entropy.
When the total Hilbert space ${\cal H}_{tot}$ is decomposed into a direct product
${\cal H}_{tot}={\cal H}_A\otimes {\cal H}_{A^c}$, we define the reduced density matrix $\rho_A$ by $\rho_A=\mbox{Tr}_{A^c}\rho_{tot}$, where $\rho_{tot}$ is the total density matrix. The entanglement entropy $S(\rho_A)$ for the subsystem $A$ is defined by
\be
S(\rho_A)=-\mbox{Tr}\rho_A\log \rho_A.
\ee

We would like to start with the definition of holographic entanglement entropy \cite{RT,HRT}
in a general setup, where we have a classical gravity dual.
In the most part of this paper, except in the last part, we assume a static gravity background in AdS/CFT and take a canonical time slice $M$. We set the total dimension of the gravitational spacetime is $d+1$ and then $M$ is the $d$ dimensional manifold. The quantum state dual to the gravity lives on the boundary $\de M$, which is in general a sum of disjoint manifolds
\be
\de M=N_1\cup N_2\cup\ddd\cup N_n.
\ee
We choose a subsystem $A$, which is also in general a sum of disjoint $d-1$ dimensional manifolds:
\be
A=A_1\cup A_2\ddd\cup A_n, \ \ \ \ A_i\subset N_i\ \ (i=1,2,\ddd,n).
\ee
We now introduce a $d-1$ dimensional surface $\Gamma_A\subset M$ such that
$\de \Gamma_A=\de A$ with the condition that $\Gamma_A$ is homologous to $A$.
Note that $\Gamma_A$ also in general consists of disjoint manifolds.
There are infinitely many candidates of $\Gamma_A$ but we choose the particular one on which the area is minimized, denoted as $\Gamma^{min}_A$. The holographic entanglement entropy \cite{RT} is given by
\be
S(\rho_A)=\frac{A(\Gamma^{min}_A)}{4G_N}, \label{hee}
\ee
where $A(\Gamma)$ represents the area of a given surface $\Gamma$.

\section{3. Entanglement Wedge Cross Section}

Let us first assume a static classical gravity dual and take a ($d$ dimensional) canonical time slice $M$. We take two subsystems $A$ and $B$ on the boundary $\de M=N$, so  that $A$ and $B$ does not have any overlap with non-zero size. In this setup we can consider the holographic entanglement entropy for $A$, $B$ and $AB(\equiv A\cup B)$ following (\ref{hee}), which
is given by the area of minimal surfaces $\Gamma^{min}_A$, $\Gamma^{min}_B$ and $\Gamma^{min}_{AB}$.

\begin{figure}
  \centering
  \includegraphics[width=3.5cm]{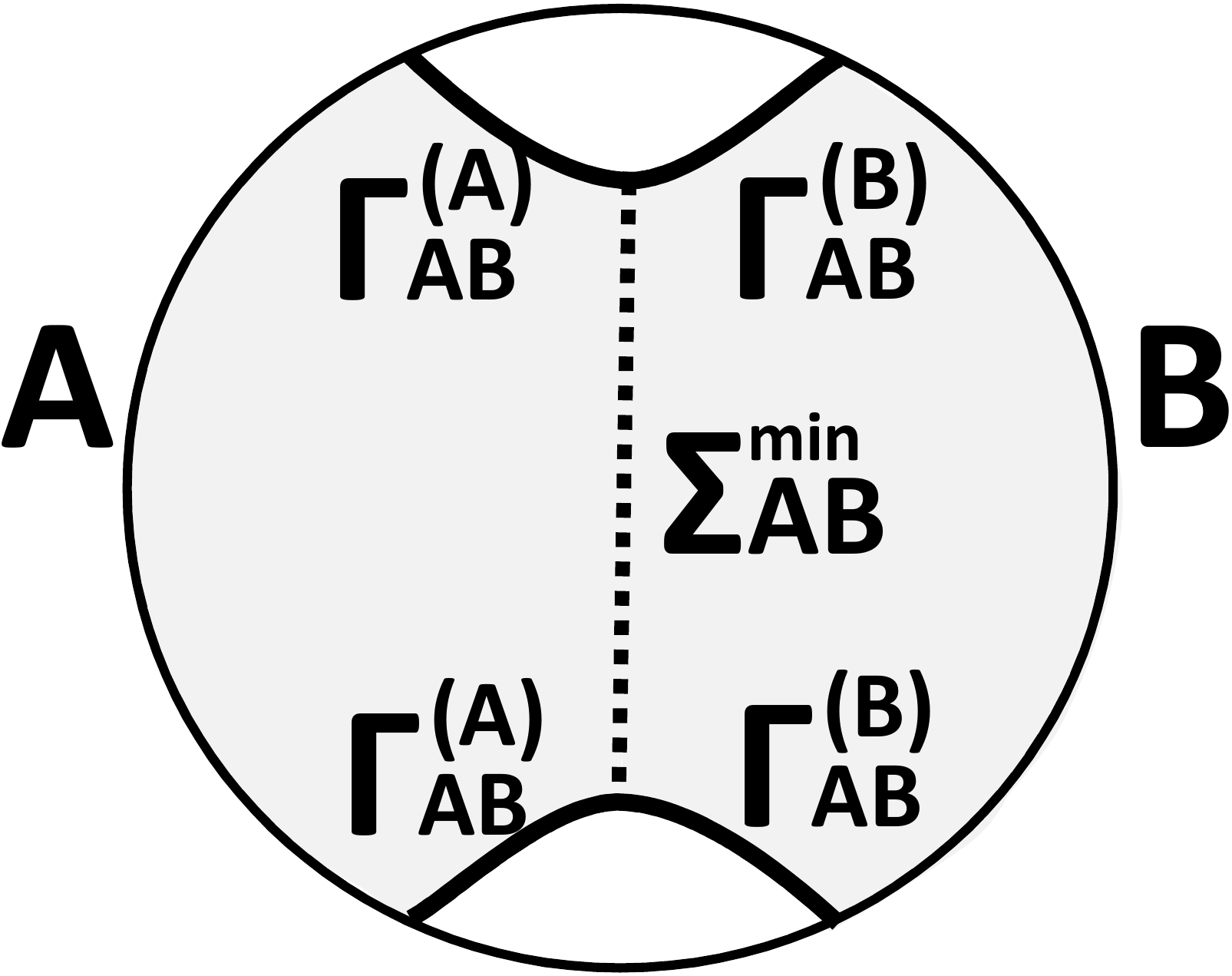}\hspace{14pt}
  \includegraphics[width=3.5cm]{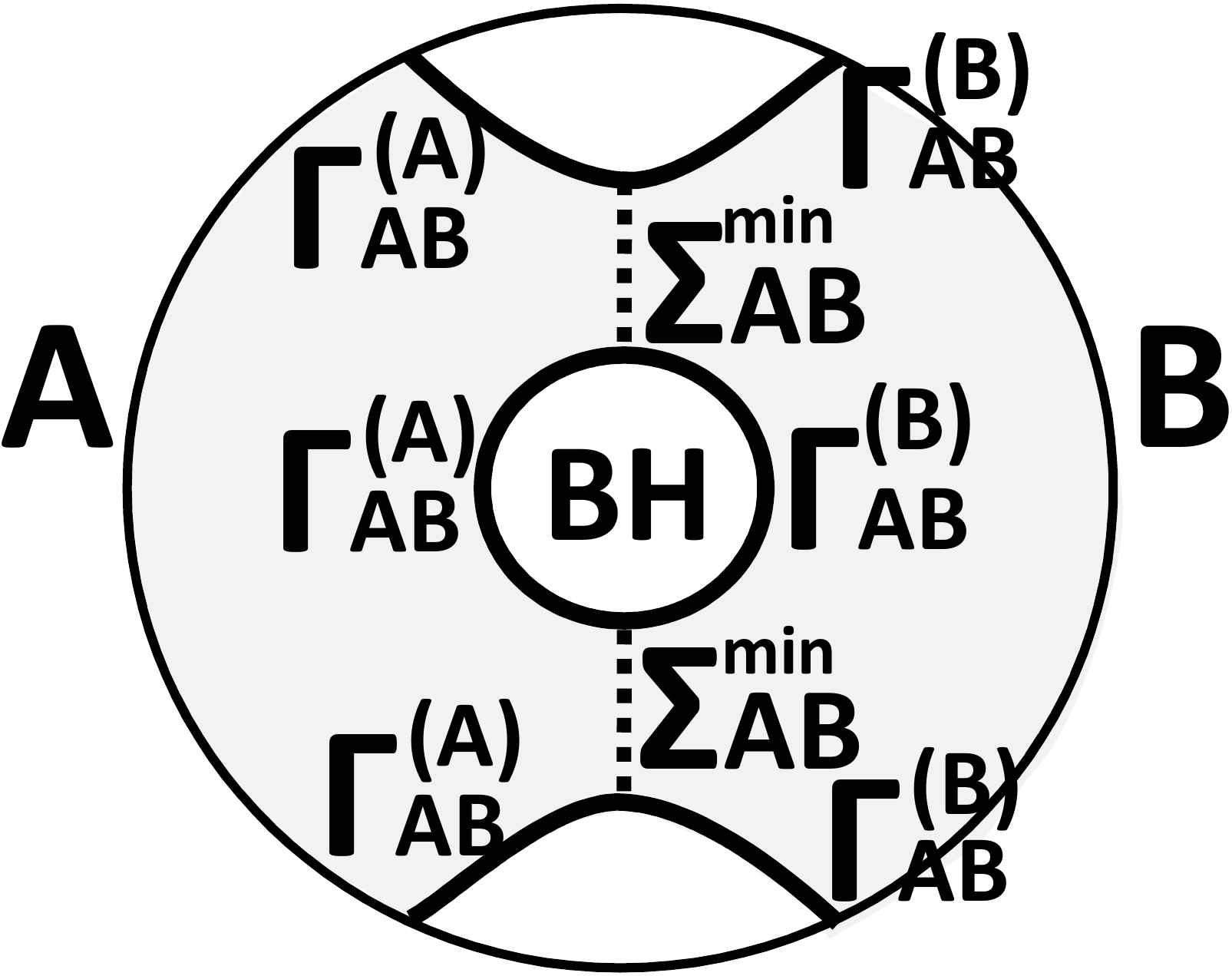}
  \caption{The gray regions are the entanglement wedges $M_{AB}$ dual to $\rho_{AB}$.
  The left one is for subsystems $A$ and $B$ in a pure state e.g. a vacuum state in a CFT.
  The right one is for subsystems for a thermal state of a CFT dual to a AdS black hole.
The surface which divides $M_{AB}$ into two parts each of which ends
  on $A$ and $B$ is defined as $\Sigma_{AB}$, which is depicted as the dotted surface.
   Equally, $\Sigma_{AB}$ is the minimal surface
  which computes the entanglement entropy between $A\cup \Gamma^{(A)}_{AB}$
  and $B\cup \Gamma^{(B)}_{AB}$.
The surface  $\Sigma^{min}_{AB}$ is obtained by minimizing the area of $\Sigma_{AB}$ by
varying the choice of $\Gamma_A$. Note also that when $A$ and $B$ gets smaller and more separated, the entanglement wedge gets disconnected into two parts in which case $\Sigma^{min}_{AB}$ becomes empty and we have $E_W=0$.}
\label{fig:EWC}
  \end{figure}

The entanglement wedge $M_{AB}$ is defined by a ($d$ dimensional) region surrounded by $A$, $B$ and $\Gamma^{min}_{AB}$ (refer to the shaded region in Fig.\ref{fig:EWC}):
\be
\de M_{AB}(\equiv N_{AB})=A\cup B\cup \Gamma^{min}_{AB}. \label{decom}
\ee
When the sizes of $A$ and $B$ are small with an enough separation, $M_{AB}$ gets disconnected into
two pieces because $\Gamma^{min}_{AB}$ also becomes disconnected.
 Note that the entanglement wedge is originally defined in the full $d+1$ dimensional spacetime as
the domain of dependence of the homology surface $R_A$, where $R_A$ is a space-like surface bounded by $A$ and $\Gamma_A$ \cite{EW1,EW2,EW3}. Therefore, strictly speaking, $M_{AB}$ is its restriction to the time slice.

Now we divide $\Gamma_{AB}$ into two parts:
\be
\Gamma^{min}_{AB}=\Gamma^{(A)}_{AB}\cup \Gamma^{(B)}_{AB}. \label{dev}
\ee
Note that $\Gamma^{(A,B)}_{AB}$ are in general unions of disjoint manifolds.
Once we choose this division (\ref{dev}), we can define the holographic entanglement entropy
$S(\rho_{\ti{\Gamma}_A})$ for
\be
\ti{\Gamma}_A\equiv A\cup \Gamma^{(A)}_{AB},  \label{tga}
\ee
assuming $M_{AB}$ is the canonical time slice of a full spacetime. If we also define
$\ti{\Gamma}_B\equiv B\cup \Gamma^{(B)}_{AB}$, then $S(\rho_{\ti{\Gamma}_A})=S(\rho_{\ti{\Gamma}_B})$.
Note that here the boundary of the entanglement wedge $M_{AB}$ is divided into two parts:
\be
\de M_{AB}=\ti{\Gamma}_A\cup \ti{\Gamma}_B.
\ee

This computation of holographic entanglement entropy is performed by finding the minimal surface $\Sigma^{min}_{AB}$ which satisfies
\ba
&&(i)\ \ \  \de\Sigma^{min}_{AB}=\de \ti{\Gamma}_A=\de \ti{\Gamma}_B,\no
&&(ii)\ \ \   \mbox{$\Sigma^{min}_{AB}$ is homologous to $\ti{\Gamma}_A$
inside $M_{AB}$}.  \label{wer}
\ea

Moreover, we minimize the area of $\Sigma^{min}_{AB}$ over all possible choices of the division (\ref{dev}). In this process, of course, we fix the manifold $M_{AB}$. This defines a quantity which we call entanglement wedge cross section, written as $E_W(\rho_{AB})$ (refer to Fig.\ref{fig:EWC}):
\ba
E_W(\rho_{AB})=\min_{\Gamma^{(A)}_{AB}\subset \Gamma^{min}_{AB}}\left[\frac{A(\Sigma^{min}_{AB})}{4G_N}\right]. \label{EWdef}
\ea

In summary, $E_W(\rho_{AB})$ computes the minimal cross section of the entanglement wedge $M_{AB}$ which connects $A$ with $B$. This is obviously a natural quantity which measures
a strength of the entanglement wedge connection. Below we would like to study the properties of this quantity.

Note that more generally we can define the entanglement wedge $M_C$ for any choice of subsystem $C$, which consists of any number of disjoint manifolds on $\de M$.
A useful property, called entanglement wedge nesting, is given by \cite{EW1,EW2,EW3}:
\be
\mbox{If}\ \ \  C\subset C',\ \  \ \mbox{then}\ \  \  M_C\subset M_{C'}. \label{EWN}
\ee
We can also show (see appendix B)
\be
\mbox{If}\ \ \  C\cap C'=\emptyset,\ \  \ \mbox{then}\ \  \  M_C\cap M_{C'}=\emptyset. \label{EWO}
\ee

\section{4. Properties of $E_W$}

First of all, from the definition (\ref{EWdef}), it is clear that if the total system $\rho_{AB}$ is a pure state, then
$\Sigma^{min}_{AB}$ coincides with $\Gamma_A=\Gamma_B$. Therefore $E_W$ gets equal to the
entanglement entropy:
\be
E_W(\rho_{AB})=S(\rho_A)=S(\rho_B), \ \mbox{when}\ \ \mbox{$\rho_{AB}$ is pure.} \label{pured}
\ee

Moreover, the holographic computation of
$E_W(\rho_{AB})$ explicitly shows that $E_W$ does not include any UV divergence as long as $A$ and $B$ do not
have any overlap with each other. We can also shows the following upper bound:
\be
E_W(\rho_{AB})\leq \min \left[S(\rho_A),S(\rho_B)\right]. \label{uppd}
\ee

It is also obvious that $E_W(\rho_{AB})$ is a non-negative quantity. When $A$ and $B$ are enough far away from each other, the mutual information $I(A:B)=S(\rho_A)+S(\rho_B)-S(\rho_{AB})$ is vanishing in the classical gravity limit \cite{He}. In this case the entanglement wedge $M_{AB}$ is disconnected and therefore $E_W(\rho_{AB})=0$. Note that the fact $I(A,B)=0$ is equivalent to $\rho_{AB}=\rho_A\otimes \rho_B$. As soon as we pass the phase transition point and $A$ gets closer to $B$, we obtain a connected entanglement wedge and have $I(A,B)>0$. In this process, $E_W(\rho_{AB})$ suddenly increases to a finite value. However we have to note that even if we have ``$I(A,B)=0$'' in the classical gravity dual computation, this just means that there is no $O(N^2)$ contribution to
$I(A:B)$, where $N$ is the gauge group rank of the dual CFT. Thus near the phase transition point we actually have $I(A:B)=O(1)$.

\begin{figure}
  \centering
  \includegraphics[width=3.5cm]{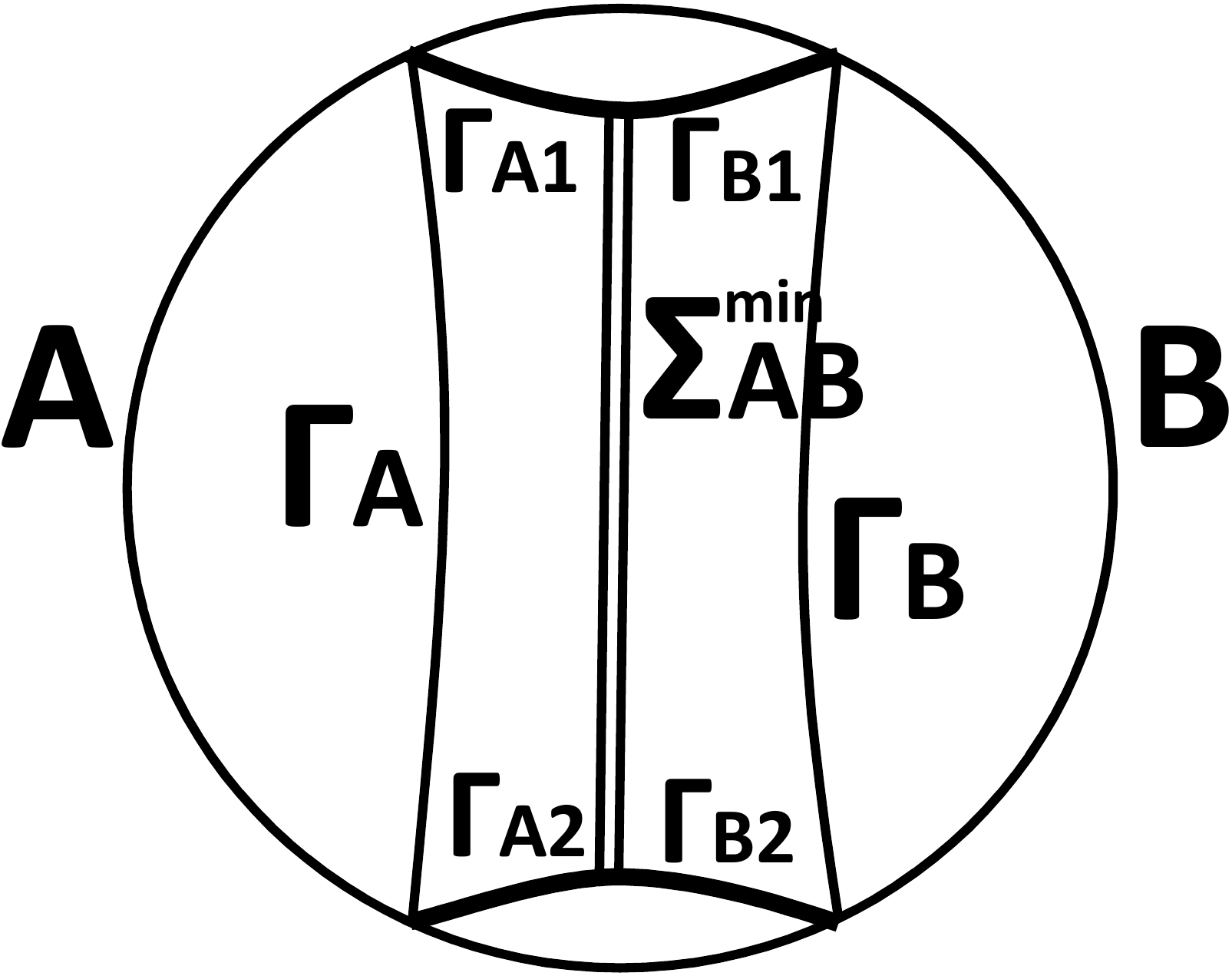}\hspace{14pt}
  \includegraphics[width=3.5cm]{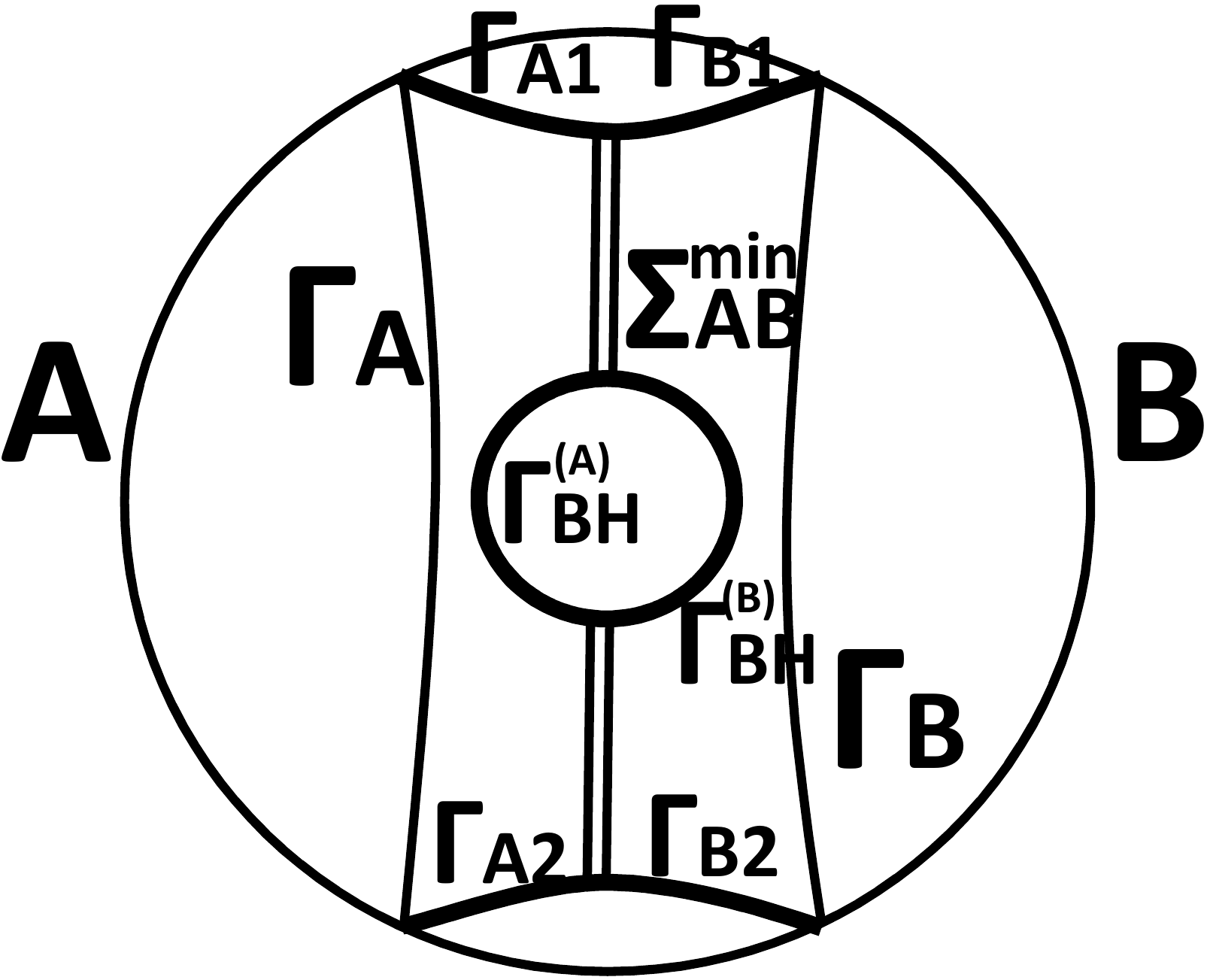}
  \caption{The proof of a bound for entanglement wedge cross section. The left picture corresponds to the case where the total system is a pure state, while the right one to the thermal state.
  It is geometrically clear that we have $S(\rho_A)+S(\rho_B)\leq 2 E_W(\rho_{AB})
  +S(\rho_{AB})$. To see this, e.g. in the right picture for a thermal state,
  we find $E_W(\rho_{AB})=A(\Sigma_{AB})$, $S(\rho_{A,B})=A(\Gamma_{A,B})$, $S(\rho_{AB})=A(\Gamma_{A1})+A(\Gamma_{A2})
  +A(\Gamma_{B1})+A(\Gamma_{B2})+A(\Gamma_{BH})$, where we set $4G_N=1$. The bound
  follows from the inequality
  $A(\Gamma_A)\leq A(\Gamma_{A1})+A(\Gamma_{A2})+A(\Sigma^{min}_{AB})+A(\Gamma^{(A)}_{BH})$ and
  a similar one for $B$. Note that $\Gamma^{(A)}_{BH}\cup \Gamma^{(B)}_{BH}$ is the black hole horizon.} \label{fig:EWCbound}
  \end{figure}

Furthermore, as first found in \cite{FH}, we can prove the following bound
\ba
E_W(\rho_{AB})\geq \frac{1}{2}I(A:B). \label{ewb}
\ea
The proof of this inequality is sketched in Fig.\ref{fig:EWCbound}.
Note that this inequality is saturated when $AB$ is a pure state.
Even though $I(A:B)$ satisfies the monogamy $I(A:BB')\geq I(A:B)+I(A:B')$ in holographic theories \cite{HHM},
the quantity $E_W(\rho_{AB})$ does not. Instead, we can show the following inequality from the entanglement wedge nesting property (\ref{EWN}):
\be
E_W(\rho_{A(BC)})\geq E_W(\rho_{AB}), \label{extens}
\ee
which is analogous to the extensiveness of mutual information equivalent to the strong subadditivity of von Neumann entropy.

Indeed, when $\rho_{ABC}$ is a pure state, we can easily find the following polygamy inequality in our gravity duals:
\be
E_W(\rho_{AB})+E_W(\rho_{AC})\geq E_W(\rho_{A(BC)}),  \label{epineqq}
\ee
which can be easily derived geometrically. Also this
actually follows from (\ref{pured}) and (\ref{ewb}).

Finally, we can show the following inequality, which is properly called
strong superadditivity:
\be
E_W(\rho_{(A\ti{A})(B\ti{B})})\geq E_W(\rho_{AB})+E_W(\rho_{\ti{A}\ti{B}}),  \label{sadd}
\ee
as is obvious from Fig.\ref{fig:SADD}. More generally, we can derive this inequality
from (\ref{EWN}) and (\ref{EWO}) as we sketch in appendix B. In particular, the equality holds when the state is product $\rho_{(A\ti{A})(B\ti{B})}=\rho_{AB}\otimes\rho_{\ti{A}\ti{B}}$.

\begin{figure}
  \centering
  \includegraphics[width=3.2cm]{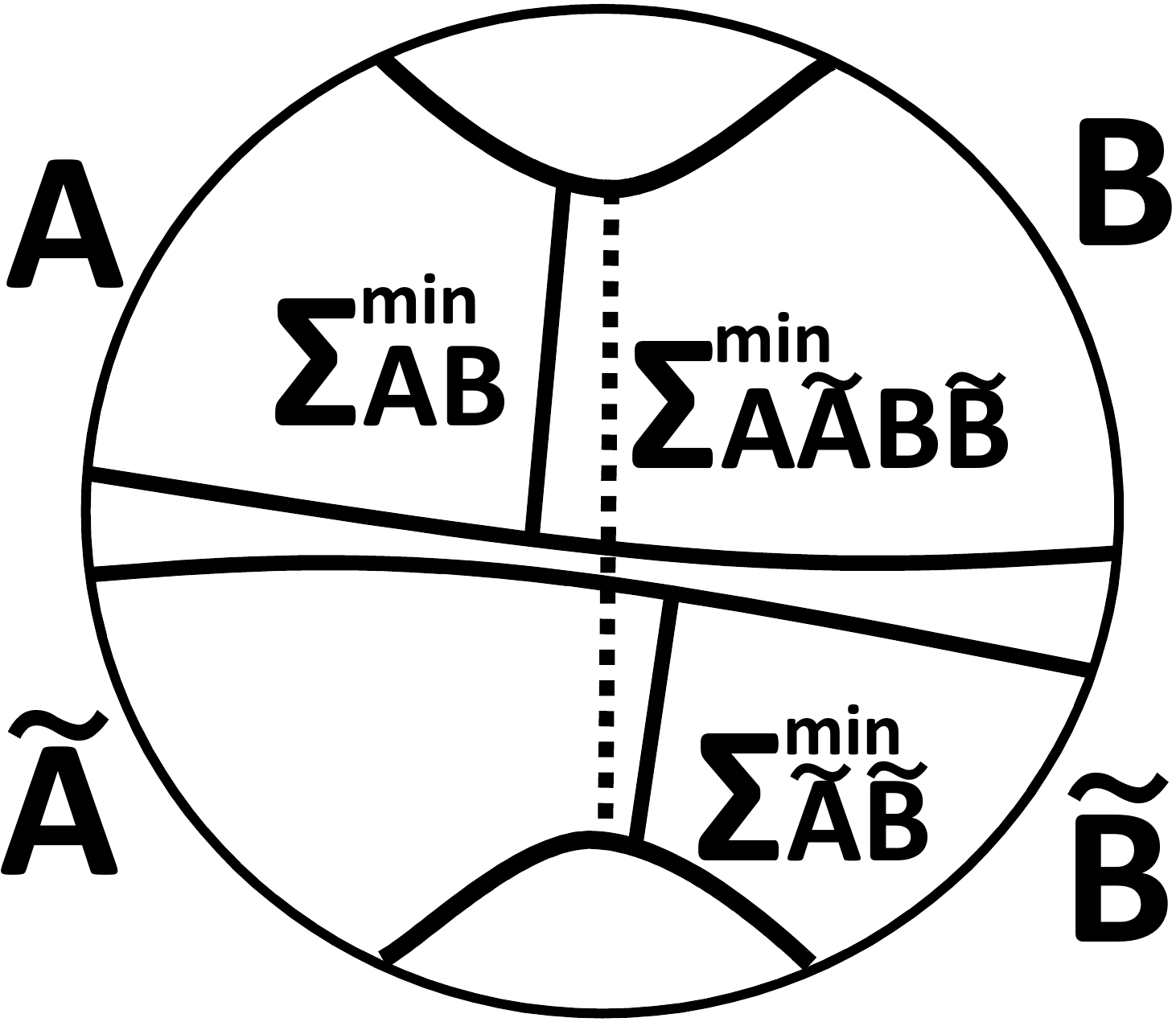}\hspace{18pt}
  \includegraphics[width=3.2cm]{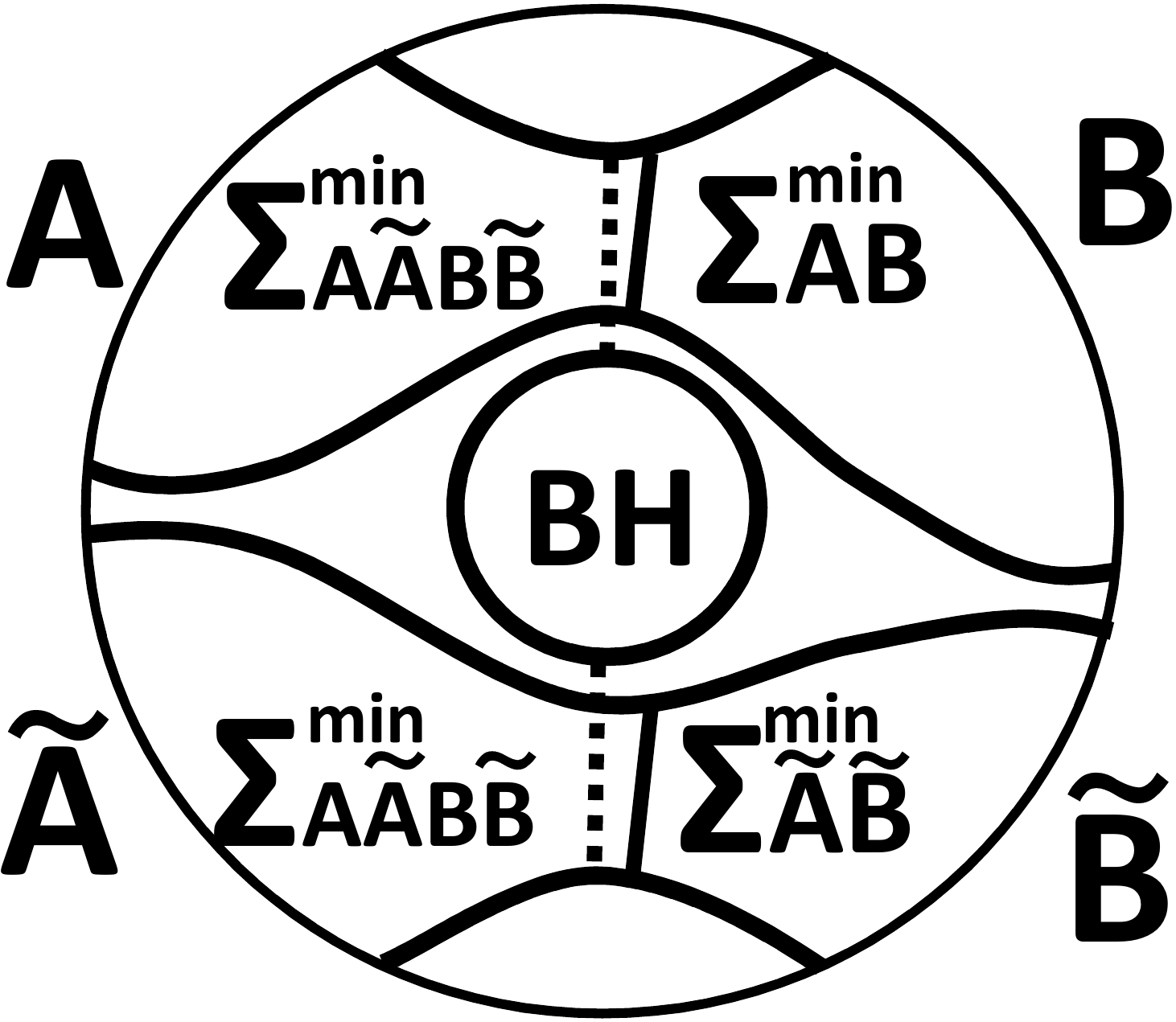}
  \caption{The proof of the strong superadditivity (\ref{sadd}).
It is obvious that the area of $\Sigma^{min}_{A\ti{A}B\ti{B}}$ is larger than the sum of
area of $\Sigma^{min}_{AB}$ and $\Sigma^{min}_{\ti{A}\ti{B}}$. $\Sigma^{min}_{AB}$
and $\Sigma^{min}_{\ti{A}\ti{B}}$ are
depicted by the thick surfaces. $\Sigma^{min}_{A\ti{A}B\ti{B}}$ is depicted as the dotted surface.}
\label{fig:SADD}
  \end{figure}

\section{5. Computations of $E_W$ in Pure AdS$_3$}

For example, as one of the simplest examples, we consider the AdS$_3/$CFT$_2$ setup and take the Poincar\'{e} coordinate. This corresponds to a vacuum state in a two dimensional holographic CFT on $\mathbb{R}^2$. The time slice is described by the metric $ds^2=\frac{dx^2+dz^2}{z^2}$,
where we set the AdS radius to be one. We choose the subsystem $A$ and $B$ to be
the interval $A=[-b,-a]$ and $B=[a,b]$ and assume $0<a<b$. To have a connected entanglement wedge, we require $I(A:B)> 0$. In this setup we evaluate as follows
($c=\frac{3}{2G_N}$ is the central charge of the dual CFT)
\ba
&& E_W(\rho_{AB})=\frac{c}{6}\log \frac{b}{a},\no
&& \frac{1}{2}I(A:B)=\frac{c}{6}\log \frac{(b-a)^2}{4ab},
\ea
from which we can explicitly confirm (\ref{uppd}) and (\ref{ewb}).

More generally, if we choose $A=[a_1,a_2]$ and $B=[b_1,b_2]$ such that $a_1<a_2<b_1<b_2$, then we obtain
\ba
&& E_W(\rho_{AB})=\frac{c}{6}\log \left(1+2z+2\s{z(z+1)}\right),\no \label{EWgen}
&& \frac{1}{2}I(A:B)=\frac{c}{6}\log z,
\ea
where $z$ is the cross ratio:
\be
z=\frac{(a_2-a_1)(b_2-b_1)}{(b_1-a_2)(b_2-a_1)},
\ee
and we assumed $z\geq 0$. Note also that since $E_W=I(A:B)=0$ for $z\leq 1$, there is a discontinuity
$\Delta E_W=\frac{c}{6}\log(3+2\s{2})$ at $z=1$.

\section{6. Computations of $E_W$ in BTZ}

Next we turn to a finite temperature state in a two dimensional holographic CFT defined on an infinite line. This corresponds to a planar BTZ black hole via AdS/CFT. The metric is given by
\ba
&& ds^2=z^{-2}\left(-f(z)dt^2+dz^2/f(z)+dx^2\right), \no \label{BTZmetric}
&&\ \ \ \  f(z)\equiv1-z^2/z_H^2,
\ea
where the location of the horizon $z_H$ is related to the inverse temperature $\beta$
via $\beta=2\pi z_H$.

\begin{figure}
  \centering
  \includegraphics[width=4cm]{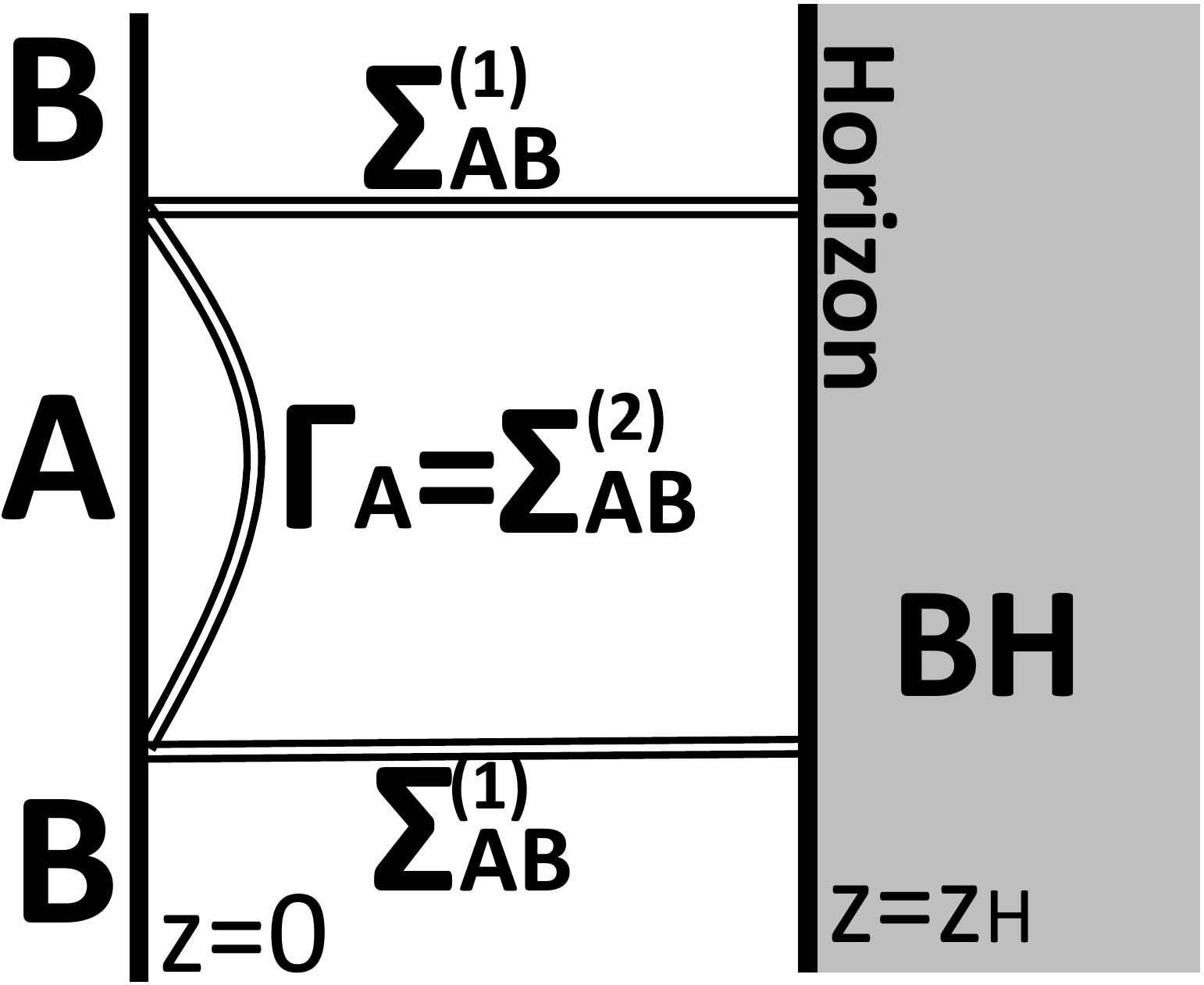}
  \caption{The computation of $E_W$ for BTZ geometry.}
\label{fig:BTZ1}
\end{figure}

We define the subsystem $A$ to be the interval $-l/2\leq x\leq l/2$ at a fixed time $t=0$.
The subsystem $B$ is defined as its complement. Obviously there are two
possibilities of the surface $\Sigma^{min}_{AB}$: one, called $\Sigma^{(1)}_{AB}$, is the union of two intervals $\ep<z\leq z_H$ at $x=l/2$ and $x=-l/2$, where $\ep$ is the UV cutoff; the other one, called $\Sigma^{(2)}_{AB}$, is the minimal surface $\Gamma_A$ (refer to Fig.\ref{fig:BTZ1}).
In the end we find
\be
E_W(\rho_{AB})=\frac{c}{3}\min\left[A^{(1)},A^{(2)}\right],
\ee
where \ba
&& A^{(1)}=\log\frac{\beta}{\pi\ep},\no
&& A^{(2)}=\log \frac{\beta\sinh\left(\frac{\pi l}{\beta}\right)}{\pi\ep}.
\ea

Therefore for $l>\beta\log(\s{2}+1)/\pi$ the disjoint surface $\Sigma^{(1)}_{AB}$ is favored, while for $l<\beta\log(\s{2}+1)/\pi$, the connected one $\Sigma^{(2)}_{AB}$ is favored. It is intriguing to note that when $l$ is vary large, the extensive contribution typical for the entanglement entropy $S(\rho_A)$, is missing in the quantity $E_W(\rho_{AB})$. Refer to appendix C
for more general computations of $E_W$ for BTZ black holes.

\section{7. Interpretation of $E_W$}

Now we would like consider how we can interpret the quantity $E_W(\rho_{AB})$ in terms of
CFTs. For this, we can consult with the whole list of correlation measures for a given mixed state $\rho_{AB}$ known in quantum information theory (refer to e.g. the excellent reviews \cite{HHHH,Book}; for a brief summary, see the appendix A of this paper). In the end, we find that the quantity called entanglement of purification $E_P(\rho_{AB})$, first introduced in \cite{EP}, behaves in the same way as the quantity $E_W(\rho_{AB})$ does. This quantity is not exactly a genuine entanglement measure as it is not always vanishing for separable states, while it is monotonic under local operations (LO), but not under classical communication (not CC). Nevertheless it is a measure of correlations between two subsystems $A$ and $B$ including classical ones.

The entanglement of purification is defined by
\be
E_P(\rho_{AB})=\min_{\rho_{AB}=\mbox{Tr}_{A'B'}|\psi\lb\la\psi|}S(\rho_{AA'}),
\label{eopdef}
\ee
where we defined $\rho_{AA'}=\mbox{Tr}_{BB'}[|\psi\lb\la\psi|]$.
The minimization in (\ref{eopdef}) is taken over any pure states $|\psi\lb\in {\cal H}_{AA'}\otimes {\cal H}_{BB'}$ ($A'$ and $B'$ are
arbitrary), which satisfy the condition $\rho_{AB}=\mbox{Tr}_{A'B'}|\psi\lb\la\psi|$. Such states $|\psi\lb$ are called purifications of $\rho_{AB}$. This quantity $E_P$ coincides with entanglement entropy for pure states as the holographic quantity $E_W$ does.

It is also useful to define its regularized version written as $E_{LOq}(\rho_{AB})$
\be
E_{LOq}(\rho_{AB})=E^\infty_{P}(\rho_{AB})\equiv
\lim_{n\to\infty}\frac{E_P(\rho_{AB}^{\otimes n})}{n}.
\ee
This quantity has a useful operational interpretation.
In the asymptotic sense, $E_{LOq}(\rho_{AB})$ counts the number of initial EPR pairs required
to create the state $\rho_{AB}$ by local operations and asymptotically vanishing communication \cite{EP}.

Indeed, the following inequalities have been shown in \cite{EP,BP}:
\ba
&&  \frac{1}{2}I(A:B)\leq E_P(\rho_{AB})\leq \min[S(\rho_A),S(\rho_B)],\label{eqa} \\
&& E_P(\rho_{A(BC)})\geq E_P(\rho_{AB}), \label{eqb} \\
&& E_P(\rho_{A(BC)})\geq \frac{1}{2}I(A:B)+\frac{1}{2}I(A:C). \label{epinw}
\ea
The first relation (\ref{eqa}) agrees with (\ref{uppd}) and (\ref{ewb}).
The second (\ref{eqb}) coincides with (\ref{extens}). The final inequality (\ref{epinw})
follows from (\ref{ewb}) by using the monogamy of mutual information in holographic CFTs
\cite{HHM}. Also if $\rho_{ABC}$ is a pure state, a polygamy relation is known \cite{BP} and this is precisely the same as (\ref{epineqq}). In this way we can confirm that all known properties agree with those for $E_W$.

From all these observations, we are tempting to conjecture
\be
E_W(\rho_{AB})=E_P(\rho_{AB}), \label{indeng}
\ee
for any holographic CFTs in the leading order of the large $N$ limit. This motivates us to call $E_W$ holographic entanglement of purification.

It is known that $E_P$ is subadditive: $E_P(\rho\otimes\sigma)\leq E_P(\rho)+E_P(\sigma)$ for a tensor product of density matrices, and the equality holds if and only if the optimal purification of $\rho\otimes\sigma$ is given by the tensor product of optimal purifications of $\rho$ and $\sigma$ (up to a local unitary equivalence) \cite{BP}. There is a numerical evidence
that the inequality is not saturated in general \cite{Wint}. In holographic CFTs, however, we expect that $E_P$ satisfies the additivity for a tensor product of density matrices and that $E_P=E^\infty_P=E_{LOq}$. This is because a holographic state is described by a single classical geometry owing to the standard saddle point approximation in gravity and a direct product state $\rho\otimes\sigma$ corresponds to two independent spacetimes, which
 clearly matches with the condition when $E_{P}$ becomes additive.

We also would like to mention the property called locking effect \cite{LockCl,LockQm}. This is a phenomenon that
a correlation measure can decrease its value by a very large amount for partially tracing out a few qubits. Indeed, the entanglement of purification $E_P$ is known to have this property
\cite{Lock}. As we have mentioned, $E_W$ has a discontinuity at the phase transition point of the entanglement wedge, which might look similar to the locking phenomena.

It is also useful to note that the lower bound in (\ref{eqa}) and the holographic counterpart (\ref{ewb}) show that $E_P$ and $E_W$ are always larger than the quantity called squashed entanglement $E_{sq}(\rho_{AB})$, which is an excellent measure of quantum entanglement for mixed states \cite{Tu,CW} and which is always smaller than $I(A:B)/2$. Refer also to \cite{HHM} for a relevance of this quantity in holographic contexts.

Now we give a heuristic explanation for why the identification (\ref{indeng}) is plausible by assuming a tensor network description. Consider a CFT state $|\psi\lb_{CFT}$ which has a classical gravity dual. We consider the reduced density matrix
$\rho_{AB}=\mbox{Tr}_{C}\left[|\psi\lb_{CFT}\la\psi|_{CFT}\right]$, where $C$ is the complement of $AB$. As defined in (\ref{EWdef}), the holographic computation of $E_W(\rho_{AB})$ is the minimum of von Neumann entropy $S(\rho_{\ti{\Gamma}_A})\left(=S(\rho_{\ti{\Gamma}_B})\right)$ when we vary the choices of $\Gamma^{(A)}_{AB}$ and $\Gamma^{(B)}_{AB}$ with the conditions (\ref{dev}) and
(\ref{tga}) as in Fig.\ref{fig:EWC}.

Indeed, this procedure of computing $E_W$ agrees perfectly with the definition of entanglement of purification (\ref{eopdef}) as we will explain below. Let us regard a time slice of AdS as a tensor network that describes a quantum state $|\psi\lb_{CFT}$ by
following \cite{MERA,Swingle,HAPPY,HQ}. In a gravity background with a tensor network description, we can define a pure state for any codimension two convex surface, called surface/state correspondence \cite{MT}. A continuous counterpart of tensor networks which describe the correct ground states in CFTs has also been formulated recently based on optimizations of path-integrals \cite{MTW,CMNTW}.  In this path integral approach, the space metric is modified by the optimization procedure and eventually we obtain a geometry of time slice of AdS from CFTs. This confirms the tensor network picture of surface/state correspondence for genuine holographic CFTs. In this correspondence, the pure state $|\psi\lb_{EW}$ realized at the boundary of the entanglement wedge $\de M_{AB}$ in the tensor network satisfies:
\be
\rho_{AB}=\mbox{Tr}_{C}\left[|\psi\lb_{CFT}\la\psi|_{CFT}\right]
=\mbox{Tr}_{\Gamma^{min}_{AB}}\left[|\psi\lb_{EW}\la\psi|_{EW}\right],
\ee
which is because $|\psi\lb_{EW}$ is obtained from $|\psi\lb_{CFT}$ by an isometry transformation. Then let us consider the minimization
of entanglement entropy in (\ref{eopdef}) with respect to the choice of
a quantum state $|\psi\lb$. It is now obvious from the geometry of tensor network
that the minimum in (\ref{eopdef}) is realized for the quantum state $|\psi\lb_{EW}$ by choosing $A'=\Gamma^{(A)}_{AB}$ and $B'=\Gamma^{(B)}_{AB}$ such that $\Sigma_{AB}$ becomes the minimum area surface with the condition (\ref{wer}) as in Fig.\ref{fig:EWC}.
Note that here we assumed that the minimized state sits within a class of states described by classical gravity duals. 
In this way, we find that the identification (\ref{indeng}) is naturally obtained in the tensor network description.

\section{8. Time-dependent Case}

 Before we finish, we would like to mention generalization of the entanglement wedge cross section $E_W$ to general time-dependent backgrounds.  For this, we do not restrict to a time slice but consider the full $d+1$ dimensional spacetime. Consider a subsystem $A$ in a CFT as in the static case. The bulk codimension two surface $\Gamma_A$ is again introduced with the conditions: (a) $\de\Gamma_A=\de A$, and (b) $\Gamma_A$ is homologous to $A$ on a codimenion one spacelike surface (i.e. a time slice) in the full Lorentzian spacetime. In this covariant setup, the holographic entanglement entropy \cite{HRT} is given by
\be
S(\rho_A)=\min_{\Gamma^{ext}_A} \left[\frac{A(\Gamma^{ext}_A)}{4G_N}\right], \label{hrt}
\ee
where $\Gamma^{ext}_A$ represents an extremal surface and the minimization is taken if there are more than one extremal surfaces.

Accordingly, we would like to define the entanglement cross section $E_W$ for general time-dependent backgrounds. We can again consider the union $N_{AB}=A\cup B\cup \Gamma^{ext}_{AB}$ as in (\ref{decom}), where $\Gamma^{ext}_{AB}$ is the extremal surface which computes the holographic entanglement entropy for the subsystem $AB=A\cup B$ following (\ref{hrt}). Note that here we do not need to specify a manifold $M_{AB}$ such that $\de M_{AB}=N_{AB}$, though $N_{AB}$ is uniquely fixed  in our present case. As before we divide $\Gamma^{ext}_{AB}$ into two parts $\Gamma^{ext}_{AB}=\Gamma^{(A)}_{AB}\cup \Gamma^{(B)}_{AB}$ and then we define
$\ti{\Gamma}_{A,B}$ as in (\ref{tga}). Next we compute the holographic entanglement entropy
for the subsystem $\ti{\Gamma}_{A}$, which is given by the area of extremal surface
$\Sigma^{ext}_{AB}$ that satisfies the previous conditions (\ref{wer}). Finally the entanglement wedge cross section is defined by minimizing w.r.t the division:
\ba
E_W(\rho_{AB})=\min_{\Gamma^{(A)}_{AB}\subset \Gamma^{ext}_{AB}}\left[\frac{A(\Sigma^{ext}_{AB})}{4G_N}\right].
\ea

We can confirm all properties which we described previously for the above covariant version. The derivations of relations (\ref{pured}) and (\ref{uppd}) are obvious. The inequalities
 (\ref{ewb}), (\ref{sadd}), (\ref{extens}) and (\ref{epineqq}) can be proved in a way very similar to the proof of strong subadditivity in the covariant setup done in \cite{EW2}.

\section{9. Conclusions}

In this paper, we considered the quantity $E_W$ defined as the minimal cross section of entanglement wedge in AdS/CFT. We observed that its properties actually coincide with those of the quantity called entanglement of purification $E_P$, which measures correlation between two subsystems for a mixed state. We conjectured that $E_W$ coincides with $E_P$ in holographic CFTs and gave a heuristic argument for this identification based on a tensor network interpretation
of AdS/CFT. It will be an important future problem to verify this conjecture by developing explicit computations in CFTs. Since this quantity has a nice operational interpretation in quantum information theory, we expect our present work will be helpful to understand operational aspects on how the AdS/CFT correspondence works.\\

{\bf Acknowledgements} We thank Veronika Hubeny, Naotaka Kubo, Nima Lashkari, Tokiro Numasawa, Hirosi Ooguri, John Preskill, Mukund Rangamani, Noburo Shiba, Tomonori Ugajin, and Guifre Vidal for useful conversations. We are very grateful to Jonathan Oppenheim for valuable comments on the properties of entanglement of purification. We would also like to thank Matt Headrick and Henry Maxfield very much for helpful comments on the draft of this paper, from which we learned that they had independent ideas on the entanglement wedge cross section from different perspectives. TT is supported by the Simons Foundation through the ``It from Qubit'' collaboration and by JSPS Grant-in-Aid for Scientific Research (A) No.16H02182. TT is also supported by World Premier International Research Center Initiative (WPI Initiative) from the Japan Ministry of Education, Culture, Sports, Science and Technology (MEXT).

\section{Appendix A: A Brief Review of Entanglement Measures}

In this appendix we give a brief review of various entanglement measures for mixed states. For detailed reviews refer to \cite{HHHH,Book,Zoo,IntroEM}.

An entanglement measure $E_{\#}(\rho_{AB})$ of quantum entanglement
between $A$ and $B$ for a given bipartite state $\rho_{AB}$ is expected to
satisfy the following conditions:\\

(a) It is non-negative and vanishing for separable states.

(b) It coincides with the entanglement entropy $S(\rho_{A})=S(\rho_{B})$
when $\rho_{AB}$ is pure.

(c) It is monotonically decreasing under local operations and classical communication (LOCC). More precisely,
if we perform LOCC on $\rho_{AB}$ and obtain the ensemble $\{p_{i},\rho_{AB}^{i}\}$,
then $E_{\#}(\rho_{AB})\geq\sum_{i}p_{i}E_{\#}(\rho_{AB}^{i})$.

(d) It is asymptotic continuous: for any states $\rho_{n},\ \sigma_{n}$
acting on $d_{n}$ dimensional Hilbert space, it follows in the asymptotic regime $n\to\infty$ that
\begin{equation}
||\rho_{n}-\sigma_{n}||_{1}\to0,\ {\rm then}\ \ \frac{E_{\#}(\rho_{n})-E_{\#}(\sigma_{n})}{\log d_{n}}\to0.
\end{equation}

(e) It is convex under classically mixing states i.e. $E_{\#}(\lambda\rho+(1-\lambda)\sigma)\leq\lambda E_{\#}(\rho)+(1-\lambda)E_{\#}(\sigma)$ where $\lambda\in[0,1]$.\\

Moreover, an entanglement (or correlation) measure $E_{\#}(\rho_{AB})$ is called\\

(i) additive if it satisfies
\begin{equation}
E_{\#}(\rho_{AB}\otimes\sigma_{\tilde{A}\tilde{B}})=E_{\#}(\rho_{AB})+E_{\#}(\sigma_{\tilde{A}\tilde{B}}),
\end{equation}

(ii) subadditive if it satisfies
\begin{equation}
E_{\#}(\rho_{AB}\otimes\sigma_{\tilde{A}\tilde{B}})\leq E_{\#}(\rho_{AB})+E_{\#}(\sigma_{\tilde{A}\tilde{B}}),
\end{equation}

(iii) strong superadditive if it satisfies
\begin{equation}
E_{\#}(\rho_{(A\tilde{A})(B\tilde{B})})\geq E_{\#}(\rho_{AB})+E_{\#}(\rho_{\tilde{A}\tilde{B}}),
\end{equation}

for any states, respectively.\\

It is known that for any normalizable measure (i.e. for
a $d$ dimensional maximally entangled state $\Phi_{d}^{+}$, we have $E_{\#}(\Phi_{d}^{+})=\log d$) which satisfies (c) and (d), the regularization $E_{\#}^{\infty}(\rho)=\lim_{n\to\infty}E_{\#}(\rho^{\otimes n})/n$
is always bounded from below by the distillable entanglement $E_{D}(\rho_{AB})$ \cite{BDSW,ED}
and from above by the entanglement cost $E_{C}(\rho_{AB})$ \cite{BDSW,Hay}:
\begin{equation}
E_{D}(\rho_{AB})\leq E_{\#}^{\infty}(\rho_{AB})\leq E_{C}(\rho_{AB}).\label{eqEDECbounds}
\end{equation}
The distillable entanglement is defined by
\begin{align}
 & E_{D}(\rho_{AB})\nonumber \\
 & =\sup_{r}\left\{ r\left|\lim_{n\to\infty}\left[\inf_{\Lambda\in{\rm LOCC}}D_{tr}\left(\Lambda(\rho_{AB}^{\otimes n}),\Phi_{2^{rn}}^{+}\right)\right]=0\right.\right\} ,
\end{align}
and the entanglement cost is defined by
\begin{align}
 & E_{C}(\rho_{AB})\nonumber \\
 & =\inf_{r}\left\{ r\left|\lim_{n\to\infty}\left[\inf_{\Lambda\in{\rm LOCC}}D_{tr}\left(\rho_{AB}^{\otimes n},\Lambda(\Phi_{2^{rn}}^{+})\right)\right]=0\right.\right\} ,
\end{align}
where $D_{tr}(\rho,\sigma)$ is the trace distance (also refer to \cite{IntroEM}). These quantities have clear operational interpretations: $E_D(\rho_{AB})\  (E_C(\rho_{AB}))$ represents the maximal (minimal) rate in which the EPR
pairs can be extracted from (are needed to produce) the state $\rho_{AB}$
by using LOCC in the asymptotic regime. If a measure is also extensive $E_{\#}(\rho_{AB}^{\otimes n})=nE_{\#}(\rho_{AB})$, the bounds \eqref{eqEDECbounds} are reduced to
\begin{equation}
E_{D}(\rho_{AB})\leq E_{\#}(\rho_{AB})\leq E_{C}(\rho_{AB}).
\end{equation}

When $\rho_{AB}$ is pure, we have $E_{D}(\rho_{AB})=E_{C}(\rho_{AB})=S(\rho_{A})$, 
and then there is the essentially unique measure (namely the entanglement entropy)
which satisfies desirable properties \cite{Unique}.

On the other hand, for mixed states, there are many inequivalent measures
of entanglement (including $E_{C}$ and $E_{D}$) and each of them
captures different types of quantum correlation.
One important class of such measures are constructed by the convex
roof. By taking an optimization over decomposing $\rho_{AB}$
into pure states as
\begin{align}
\rho_{AB} =\sum_{i}p_{i}|\psi_{i}\lb\la\psi_{i}|_{AB},\ \ \ p_{i} \geq0,\ \sum_{i}p_{i}=1,
\end{align}
we reach the entanglement of formation $E_{F}(\rho_{AB})$ \cite{BDSW}:
\begin{align}
 & E_{F}(\rho_{AB})\nonumber \\
 & =\inf_{\rho_{AB}=\sum_{i}p_{i}|\psi_{i}\lb\la\psi_{i}|_{AB}}\sum_{i}p_{i}S({\rm Tr}_{B}|\psi_{i}\lb\la\psi_{i}|).\label{eq:eof}
\end{align}
This measure satisfies all of (a)-(e) conditions \cite{Book}. It is known that the regularized entanglement of formation $E_{F}^{\infty}$ is equal to $E_{C}$: $\lim_{n\to\infty}E_{F}(\rho^{\otimes n})/n=E_{C}(\rho)$
\cite{Hay}. $E_{F}$ is also subadditive, which immediately leads to a bound $E_{C}\leq E_{F}$.
However, $E_{F}$ and $E_{C}$ are different in general i.e. $E_{F}$
is not additive \cite{Hast}.

There is another method to find an entanglement measures. This is
based on a certain distance between a given density matrix $\rho_{AB}$
and a set of separable states. The most famous one is the relative
entropy of entanglement $E_{R}(\rho_{AB})$ \cite{REE}:
\begin{equation}
E_{R}(\rho_{AB})=\inf_{\sigma_{AB}\in\mbox{Sep.}}S(\rho_{AB}||\sigma_{AB}),
\end{equation}
where $S(\rho||\sigma)={\rm Tr}\left(\rho\log\rho-\rho\log\sigma\right)$
is the relative entropy. It also belongs to the good measure class, but is not additive (extensive): $E_{R}^{\infty}\neq E_{R}$ \cite{Book}. The following
inequalities instead has been shown: $E_{D}\leq E_{R}\leq E_{F}$ \cite{Rai,VP,HV}.
Also the inequality $E_{R}(\rho_{AB})\leq I(A:B)$ was noted in \cite{LL}.

Now let us develop the convex roof procedure a little more.
We introduce an extension of a given state $\rho_{AB}$ acting
on the Hilbert space $H_{A}\otimes H_{B}\otimes H_{C}$
such that
\begin{equation}
\mbox{Tr}_{C}\rho_{ABC}=\rho_{AB},
\end{equation}
where we can choose any $H_{C}$ and $\rho_{ABC}$ with the above
condition. This leads to the squashed entanglement $E_{sq}(\rho_{AB})$
\cite{Tu,CW} defined by
\begin{equation}
E_{sq}(\rho_{AB})=\frac{1}{2}\inf_{\rho_{AB}=\mbox{Tr}_{C}\rho_{ABC}}I(A:B|C).
\end{equation}
Here
\begin{equation}
I(A:B|C)=S(\rho_{AC})+S(\rho_{BC})-S(\rho_{C})-S(\rho_{ABC})\geq0,
\end{equation}
is the quantum conditional mutual information, which is non-negative due
to the strong subadditivity of von Neumann entropy. It is thought that $E_{sq}$ is the
most promising measure of entanglement for mixed states. First, It satisfies (a)-(e) and the additivity \cite{Book} with the bounds $E_{D}\leq E_{sq}=E_{sq}^{\infty}\leq E_{C}$. It is also faithful i.e. $E_{sq}(\rho)=0$ if and only if $\rho$ is separable \cite{faithful}.
In particular, it satisfies the monogamy relation \cite{KW}
\begin{equation}
E_{sq}(\rho_{A(BB')})\geq E_{sq}(\rho_{AB})+E_{sq}(\rho_{AB'}),
\end{equation}
which represents a prominent feature of quantum correlations in terms of shareability. From this monogamy,
we can derive the strong superadditivity
\begin{equation}
E_{sq}(\rho_{(A\tilde{A})(B\tilde{B})})\geq E_{sq}(\rho_{AB})+E_{sq}(\rho_{\tilde{A}\tilde{B}}).
\end{equation}
It is useful to note that $E_{F}$ and $E_{R}$ does
not satisfy the strong superadditivity, while $E_{D}$ does \cite{Zoo}. We can also
show the upper and lower bound in terms of mutual information \cite{CW}
\begin{equation}
\frac{1}{2}(I(A:B)-S(\rho_{AB}))\leq E_{sq}(\rho_{AB})\leq\frac{1}{2}I(A:B).\label{squp}
\end{equation}

In this way we find a general relation between the entanglement measures
\begin{align}
E_{D} & \leq E_{sq}\leq E_{C}\leq E_{F},\no
E_{D} & \leq E_{R}\leq E_{F}.
\end{align}

Finally, we mention about a quantity called entanglement of purification
$E_{P}$ introduced in \cite{EP}. It measures not an amount of entanglement,
but a total correlation between $A$ and $B$ as the mutual information $I(A:B)$
does. Properties of $E_{P}$ were reviewed in the context of this paper (see \cite{BP} for details). In addition to them a bound in
terms of the $E_{F}$ has been proven: $E_{F}\leq E_{P}$ \cite{EP}.
It would be worth noting that for some quantum states $E_{P}$ exceeds $I$.

The regularization of entanglement of purification $E_P^{\infty}$ coincides with $E_{LOq}(\rho_{AB})$ \cite{EP} defined by
\begin{align}
 & E_{LOq}(\rho_{AB})\nonumber \\
 & =\inf_{r}\left\{ r\left|\lim_{n\to\infty}\left[\inf_{\Lambda\in{\rm LOq}}D_{tr}\left(\rho_{AB}^{\otimes n},\Lambda(\Phi_{2^{rn}}^{+})\right)\right]=0\right.\right\}.
\end{align}
This quantity is analogous to $E_{C}(\rho_{AB})$ with the restriction of optimizing procedure to local operations and
asymptotically vanishing communication (LOq). It has similar operational interpretation as $E_{C}$ and a bound $E_{C}\leq E_{LOq}$ is clear by its definition.

\section{Appendix B: Proof of the Strong Superadditivity}

We can prove the strong superadditivity of $E_{W}$ in general from (\ref{EWN}) and
(\ref{EWO}). Note that (\ref{EWO}) is obvious since if there is an overlap between $M_{C}$ and $M_{C'}$ we can choose another minimal surface which gives a smaller area than that of $\Gamma_{C}^{min}$, which contradicts with the definition of $M_C$ (see also \cite{EW2}). The property (\ref{EWN}) tells us  $M_{AB}\subset M_{A\ti{A}B\ti{B}}$ and $M_{\ti{A}\ti{B}}\subset M_{A\ti{A}B\ti{B}}$. We also have $M_{AB}\cap M_{\ti{A}\ti{B}}=\emptyset$ from (\ref{EWO}). 
Thus we find $M_{AB}\cup M_{\ti{A}\ti{B}}\subset M_{A\ti{A}B\ti{B}}$ without overlap. It leads the strong superadditivity of $E_W$.

Let us consider an example in Fig.\ref{fig:SADD}, assuming the entanglement wedge between $A=[a_{1},a_{2}]$ and $\ti{B}=[b_{1},b_{2}]$ is connected i.e.~$E_{W}(\rho_{A\ti{B}})>0$. In this setup (\ref{EWO}) says that there is no connected entanglement wedge between any subsystems $\ti{A}\subset\bar{A}=[a_{2},b_{1}]$
and $B\subset\bar{B}=(A\cup\bar{A}\cup \ti{B})^{c}$. Note that this fact can be also easily seen from the mutual information between $\bar{A}$ and $\bar{B}$:
\begin{equation}
I(\bar{A}:\bar{B})=S(\rho_{A\ti{B}})-S(\rho_{A})-S(\rho_{\ti{B}})=-I(A:\ti{B})<0,
\end{equation}
and $I(\ti{A}:B)\leq I(\bar{A}:\bar{B})$. Therefore $E_{W}(\rho_{\ti{A}B})=0$ must follow. Likewise, if we assume $E_{W}(\rho_{\ti{A}B})>0$, then we have $E_{W}(\rho_{A\ti{B}})=0$. 
The strong superadditivity in these diagonal setups is now clear with (\ref{EWN}):
\begin{equation}
E_{W}(\rho_{(A\tilde{A})(B\tilde{B})})\text{\ensuremath{\ge}}E_{W}(\rho_{A\tilde{B}})+E_{W}(\rho_{\tilde{A}B}),
\end{equation}
as at least one term on the right-hand side does vanish.

If we consider other setup of subsystems e.g. with interchanging the position of $\ti{A}$ with $\ti{B}$ in Fig.\ref{fig:SADD}, the strong superadditivity can also be proven in the same manner containing disjoint $\Sigma^{min}_{(A\ti{A})(B\ti{B})}$.

\section{Appendix C: COMPUTATIONS of $E_{W}$ in ADS$_{3}$/CFT$_{2}$}

We provide the general forms of $E_{W}$ for various
setups in AdS$_{3}$/CFT$_{2}$. In this section we will set $R=4G_{N}=1$
for simplicity.
\subsection*{Pure AdS$_{3}$ }

The expression of $E_{W}$ in the Poincar\'{e} coordinate is already
obtained in (\ref{EWgen}).
In the global coordinate, which
corresponds to the vacuum state in holographic CFTs on a cylinder of circumference
$L$, one can employ the conformal map from a cylinder to a plane
and read off the transition from that of the mutual information. The result
is
\begin{equation}
E_{W}(\rho_{AB})=\log\left(1+2\tilde{z}+2\sqrt{\tilde{z}(\tilde{z}+1)}\right),
\end{equation}
where we defined
\begin{equation}
\tilde{z}\equiv\frac{\sin\left(\frac{\pi(a_{2}-a_{1})}{L}\right)\sin\left(\frac{\pi(b_{2}-b_{1})}{L}\right)}{\sin\left(\frac{\pi(b_{1}-a_{2})}{L}\right)\sin\left(\frac{\pi(b_{2}-a_{1})}{L}\right)},
\end{equation}
and $A=[a_{1},a_{2}],\ B=[b_{1},b_{2}]$.

\subsection*{BTZ black hole}

The metric is given in (\ref{BTZmetric}). The BTZ black hole is a quotient spacetime
of the pure AdS$_{3}$ and the previous result (\ref{EWgen}) can be used to get the form of $E_{W}$. Similar to the global coordinate in
pure AdS$_{3}$, we obtain
\begin{eqnarray}
E_{W}(\rho_{AB}) & = & \log\left(1+2\zeta+2\sqrt{\zeta(\zeta+1)}\right),
\end{eqnarray}
where we defined
\begin{equation}
\zeta\equiv\frac{\sinh\left(\frac{\pi(a_{2}-a_{1})}{\beta}\right)\sinh\left(\frac{\pi(b_{2}-b_{1})}{\beta}\right)}{\sinh\left(\frac{\pi(b_{1}-a_{2})}{\beta}\right)\sinh\left(\frac{\pi(b_{2}-a_{1})}{\beta}\right)}.
\end{equation}
and $A=[a_{1},a_{2}],\ B=[b_{1},b_{2}]$.

\begin{figure}
  \centering
  \includegraphics[width=2.5cm]{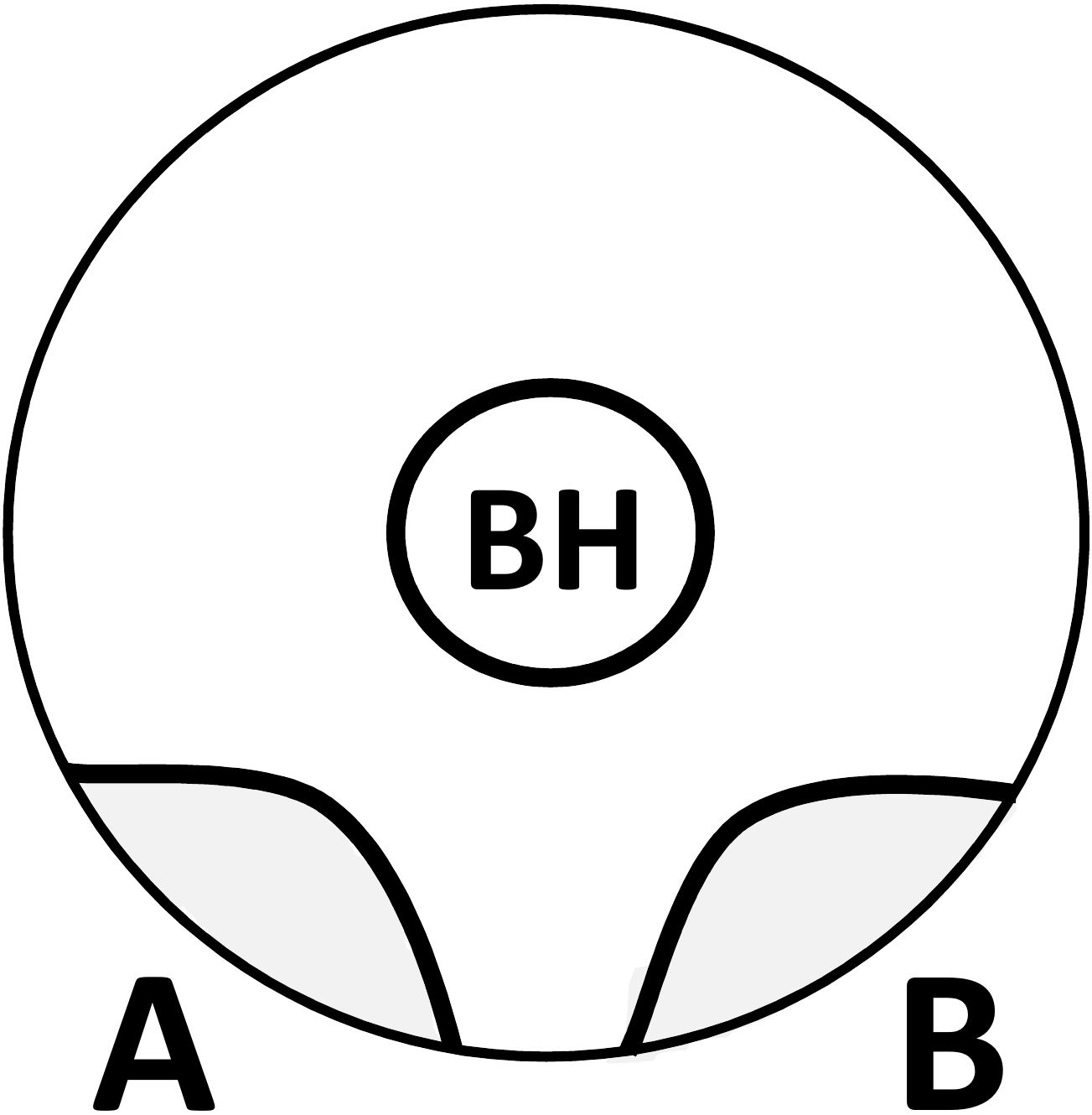}\hspace{5pt}
  \includegraphics[width=2.5cm]{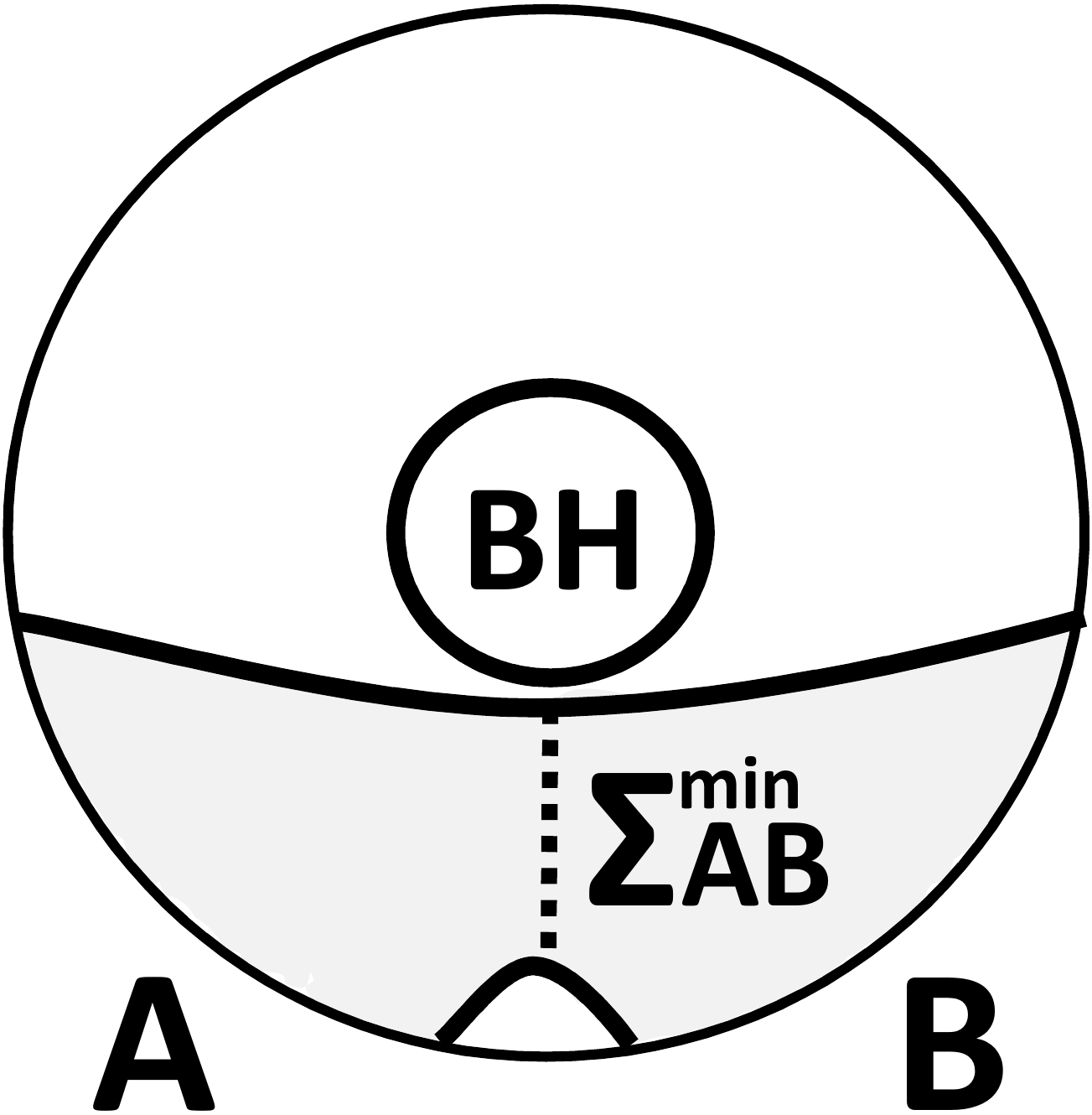}\hspace{5pt}
  \includegraphics[width=2.5cm]{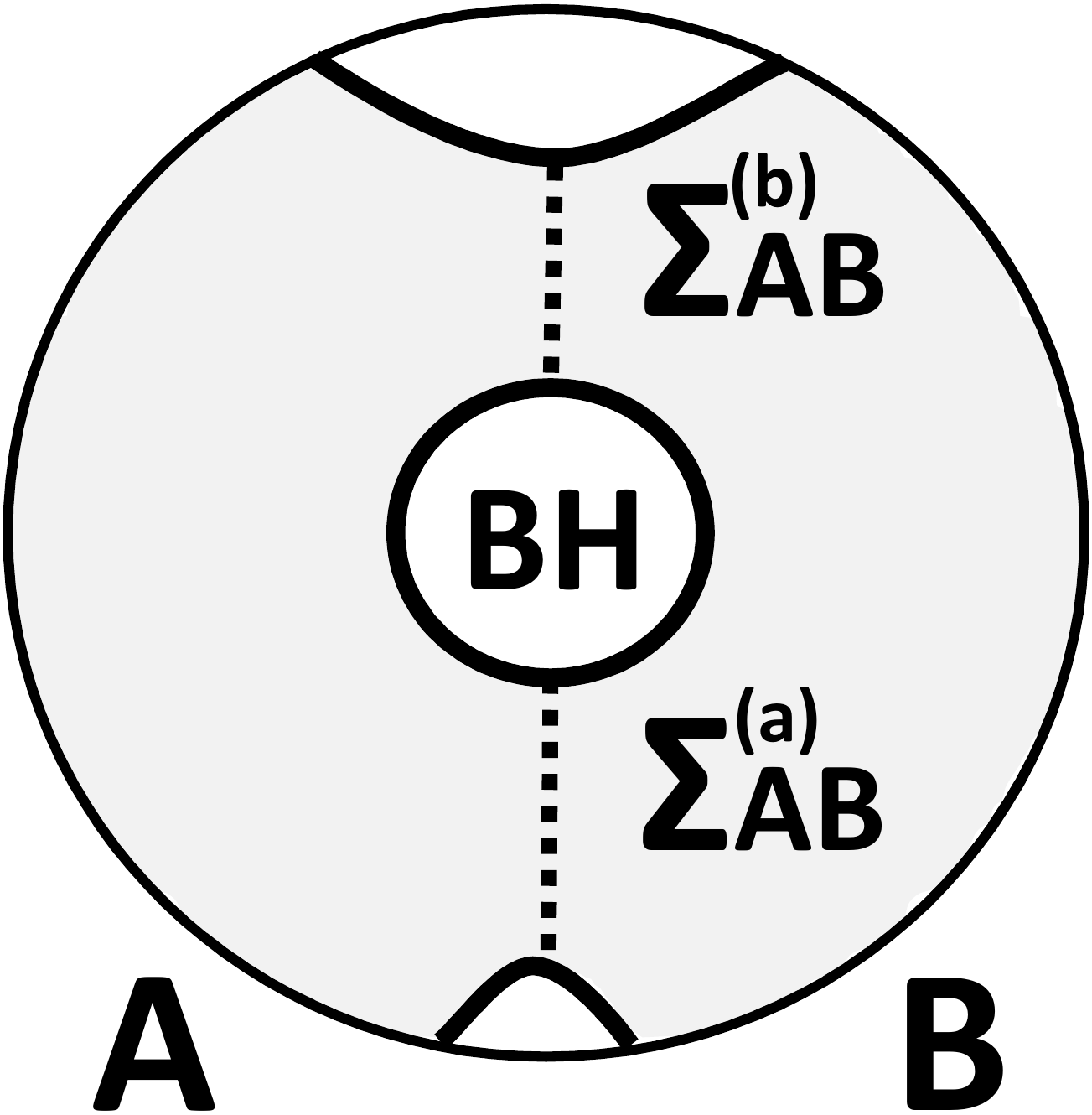}
  \caption{Three phases of the entanglement wedge for a symmetric setup in a global BTZ black hole.}
  \label{fig:BTZPT}
\end{figure}

There is another candidate for $\Sigma_{AB}^{min}$ in a global BTZ black hole when the subsystems are sufficiently large so that the entanglement wedge $M_{AB}$ surrounds the horizon. This is the disconnected codimension-2 surfaces which have endpoints on the black hole horizon, as depicted in the right picture of Fig.\ref{fig:BTZPT}.

\begin{figure}
  \centering
  \includegraphics[width=5.5cm]{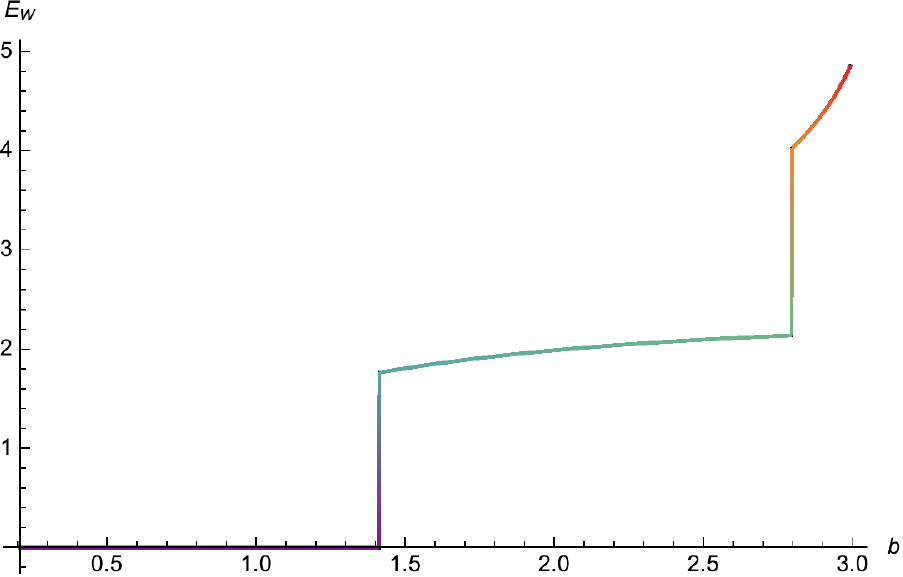}
  \caption{$E_{W}(\rho_{AB})$ is plotted as a function of $b\in(a,\pi)$ setting $L=\beta=2\pi,\ a=\pi/15$.
  The left jump happens when $I(A:B)$ vanishes and the right when $\Gamma_{AB}^{min}$
  wraps the horizon.}
  \label{fig:BTZPLOT}
\end{figure}

Let us consider a symmetric
setup $A=[-b,-a],\ B=[a,b]$ for simplicity. In this setup such disconnected surfaces anchored to the horizon are clearly favored for $\Sigma_{AB}^{min}$ when the subsystems are large enough. The area of $\Sigma_{AB}^{(a)}$ in Fig.\ref{fig:BTZPT} is given by
\begin{eqnarray}
A(\Sigma_{AB}^{{\rm (a)}}) & = & \int_{z_{*}}^{z_{H}}\frac{dz}{z\sqrt{f(z)}}=\log\left(\frac{2\cosh^{2}(\frac{\pi a}{\beta})}{\sinh(\frac{2\pi a}{\beta})}\right),
\end{eqnarray}
where the turning point of a minimal surface is given by $z_{*}=z_{H}\tanh (l/2z_{H})$ for a subsystem of size $l$. The same holds for $\Sigma_{AB}^{(b)}$ by replacing $a$ to
$L/2-b$. Thus we find
\begin{eqnarray}
E_{W}(\rho_{AB}) & = & \log\left(\frac{4\cosh^{2}(\frac{\pi a}{\beta})}{\sinh(\frac{2\pi a}{\beta})}\frac{\cosh^{2}(\frac{\pi(L/2-b)}{\beta})}{\sinh(\frac{2\pi(L/2-b)}{\beta})}\right),
\end{eqnarray}
in this phase. As the subsystems become larger, discontinuity of $E_{W}$ happens at most twice because of phase transitions
of $\Gamma_{AB}^{min}$. This is plotted in Fig.\ref{fig:BTZPLOT}

\begin{figure}
  \centering
  \includegraphics[width=3.5cm]{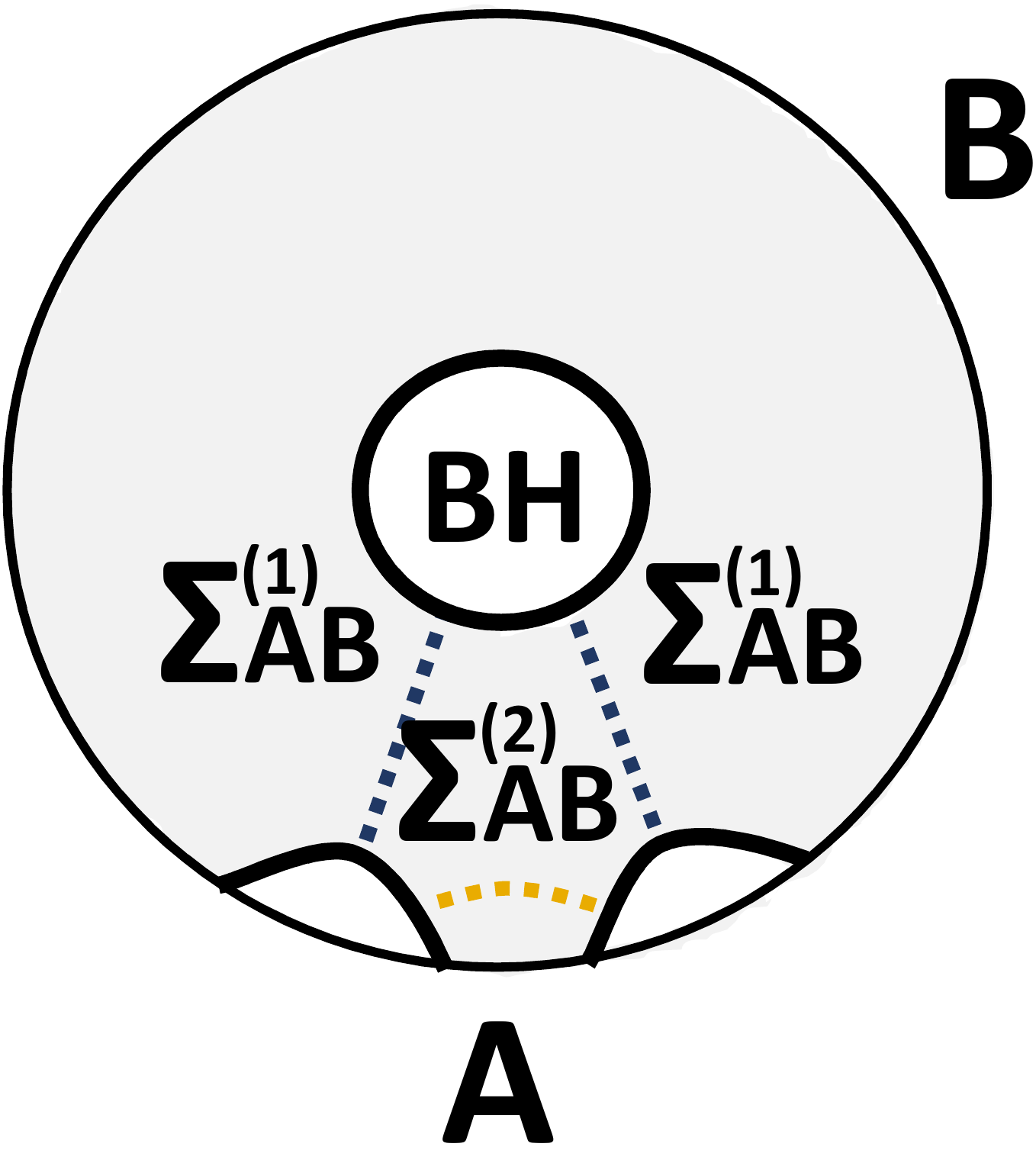}
  \caption{Two candidates exist for $\Sigma_{AB}^{min}$ in such phases.
If the subsystem $B$ becomes considerably small, a disconnected entanglement wedge will be favored.}
  \label{fig:BTZPT2}
\end{figure}

\begin{figure}
  \centering
  \includegraphics[width=5.5cm]{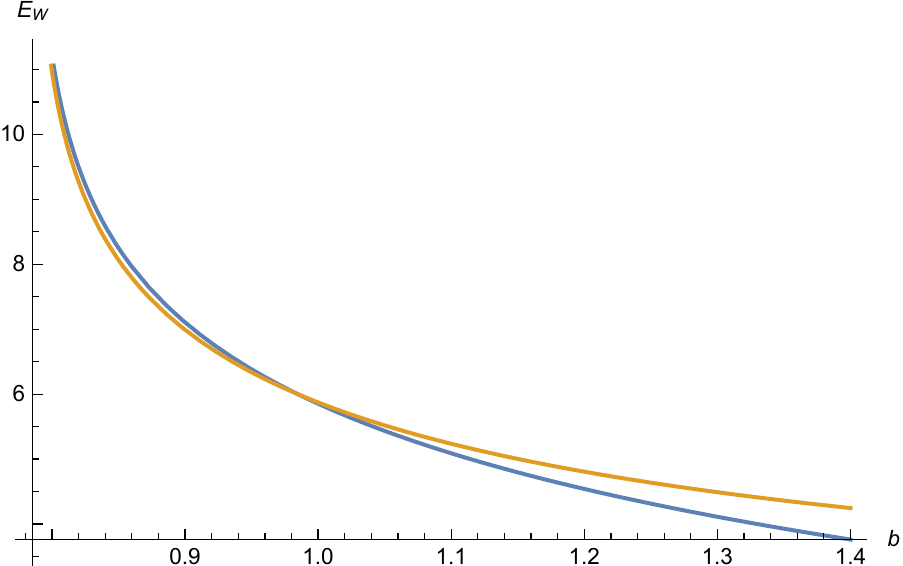}
  \caption{$E_{W}(\rho_{AB})$ is plotted as a function of $b\in(a,\pi)$ setting $L=\beta=2\pi$,
   $a=\pi/4$. Here the subsystems are $A=[-a,a],\ B=[b,L-b]$. The yellow (blue) line corresponds to the connected (disconnected) surface in Fig.\ref{fig:BTZPT2}, respectively. An entanglement wedge between $A$ and $B$ are connected in this range.}
  \label{fig:BTZPLOT2}
\end{figure}

 On the other hand, $E_{W}$ also gets a different phase
transition due to choosing the minimal candidate as depicted in Fig.\ref{fig:BTZPT2}-\ref{fig:BTZPLOT2}, though in this case the value of $E_W$ changes continuously.

\end{document}